\documentclass[aps,showpacs,superscriptaddress,groupedaddress,nofootinbib]{revtex4}  % for review and submission
\usepackage{graphicx}  % needed for figures
\usepackage{dcolumn}   % needed for some tables
\usepackage{bm}        % for math
\usepackage{amssymb}   % for math
\usepackage{amsmath,amsfonts,mathtools}
\usepackage{mathtools}
\usepackage{cases}
\usepackage{mathrsfs}
\usepackage{color}
\newcommand{\calR}{{\cal R}}
\newcommand{\be}{\begin{equation}}
\newcommand{\ee}{\end{equation}}
\newcommand{\bea}{\begin{eqnarray}}
\newcommand{\eea}{\end{eqnarray}}
\newcommand{\nn}{\nonumber}
\newcommand{\bx}{{\bm x}}
\newcommand{\bk}{{\bm k}}
\newcommand{\bl}{{\bm \ell}}
\newcommand{\bp}{{\bm p}}
\newcommand{\bq}{{\bm q}}

\usepackage[symbol]{footmisc}

\begin{document}

\begin{flushright}
    YITP-20-75\\
    IPMU20-0054
\end{flushright}
	\title{Gravitational Waves Induced by Scalar Perturbations with a Lognormal Peak}
	\author{Shi Pi${}^{a,b}$, and Misao Sasaki${}^{a,c,d}$\\
		\it
		$^{a}$ Kavli Institute for the Physics and Mathematics of the Universe (WPI), Chiba 277-8583, Japan\\
		$^{b}$ CAS Key Laboratory of Theoretical Physics, Institute of Theoretical Physics,\\
		Chinese Academy of Sciences, Beijing 100190, China \\
		$^{c}$ Yukawa Institute for Theoretical Physics, Kyoto University, Kyoto 606-8502, Japan\\
		$^{d}$ Leung Center for Cosmology and Particle Astrophysics,\\National Taiwan University, Taipei 10617}%, Taiwan}
	\date{\today}
\begin{abstract}
We study the stochastic gravitational wave (GW) background induced by the primordial scalar perturbation
with the spectrum having a lognormal peak of width $\Delta$ at $k=k_*$.
We derive an analytical formula for the GW spectrum $\Omega_\text{GW}$ for 
both narrow ($\Delta\ll1$) and broad  ($\Delta\gtrsim1$) peaks.
 In the narrow-peak case, the spectrum has a double peak feature with the sharper peak at 
 $k= 2k_*/\sqrt{3}$.  On the infrared (IR) side of the spectrum,  we find power-law behavior
 with a break at $k=k_b$ in the power-law index where it chages from $k^3$ on the far IR side to $k^2$ 
 on the near IR side.  We find the ratio of the break frequency to the peak frequency is determined by
 $\Delta$ as $f_b/f_p\approx\sqrt{3}\Delta$, where $f_b$ and $f_p$ are the break and peak
 frequencies, respectively.
  In the broad-peak case, we find the GW spectrum also has a lognormal peak at $k=k_*$
  but with  a smaller  width of $\Delta/\sqrt2$. 
  Using these derived analytic formulae, we also present expressions for 
  the maximum values of $\Omega_\text{GW}$ for both narrow and broad cases. 
  Our results will provide a  useful tool in searching for the induced GW  signals in the coming decades.
\end{abstract}
	\maketitle
	
\section{introduction}
The detection of gravitational waves (GWs) from the mergers of black holes (BHs) and neutron stars by LIGO/Virgo~\cite{LIGO}
%~\cite{Abbott:2016blz,Abbott:2016nmj,Abbott:2017vtc,Abbott:2017gyy,Abbott:2017oio,TheLIGOScientific:2017qsa,Abbott:2020uma} 
has marked the beginning of the era of gravitational wave astronomy. 
Besides GW bursts from mergers, there are also stochastic backgrounds of
various origins  such as GWs from binary inspirals, GWs from first order phase transitions in
the early universe, primordial GWs from inflation, and the induced GWs from the primordial
scalar perturbation. 
In the spatially homogeneous and isotropic background, 
the tensor  and scalar perturbations are decoupled at linear order,
but they are coupled to each other at nonlinear order~\cite{earlyGW},
%\cite{Matarrese:1992rp,Matarrese:1993zf,Matarrese:1997ay,Noh:2004bc,Carbone:2004iv,Nakamura:2004rm},
 giving rise to the induced GWs or secondary GWs~\cite{Baumann:2007zm,Ananda:2006af,Osano:2006ew,Saito:2008jc,Assadullahi:2009nf,Assadullahi:2009jc,Bugaev:2009zh,Saito:2009jt,Bugaev:2010bb,Alabidi:2012ex,Alabidi:2013wtp,Nakama:2016gzw,Gong:2017qlj,Espinosa:2018eve,Cai:2018dig,Bartolo:2018rku,Unal:2018yaa,Cai:2019amo,Inomata:2019zqy,Inomata:2019ivs,Ota:2020vfn,Kohri:2018awv,Bartolo:2018evs,Gong:2019mui,Yuan:2019udt,Inomata:2020lmk}. 
The primordial scalar perturbation (more precisely scalar-type curature perturbation) 
produces the cosmic microwave background radiation (CMB) anisotropy
 and seeds the large scale structure of the universe,
 and its properties are tightly constrained observations.
Namely, it is Gaussian with a nearly scale-invariant spectrum with the rms amplitude of $10^{-5}$ 
on scales larger than around $1~\text{Mpc}$~\cite{Aghanim:2018eyx}.
 The GWs induced by such primordial scalar perturbation is also nearly scale-invariant at frequencies
 higher than $10^{-15}$~Hz, but the amplitude is negligibly small to be detected~\cite{Baumann:2007zm},
and perhaps smaller than GWs from the primordial tensor perturbation~\cite{Watanabe:2006qe,Pi:2019ihn}.

However, on scales much smaller than $1$~Mpc ,
the amplitude of the primordial scalar perturbation is ony weakly constrained,
and hence it is possible to be large~\cite{Byrnes:2018txb,Inomata:2018epa,PRconstraint}. %{Dalianis:2018ymb,Lu:2019sti,Kalaja:2019uju,Ozsoy:2019lyy}. 
An interesting consequence associated with a large scalar perturbation is the formation of
primordial black holes (PBHs). 
If the rms scalar perturbation amplitude is large at some specific wavenumber, 
PBHs can form from the high $\sigma$ peaks when the corresponding wavelengths reenter  the Hubble 
 horizon~\cite{PBHformation1,PBHformation2}.
 %~\cite{Zeldovich:1963,Hawking:1971ei,Carr:1974nx,Meszaros:1974tb,Carr:1975qj,Khlopov:1985jw,Young:2014ana,Germani:2018jgr,Wu:2020ilx,DeLuca:2019qsy,Ezquiaga:2019ftu}. 
  The current observational constraints does not exclude the existence of a substantial amount of such PBHs.
There are actually several interesting ``mass  windows''~\cite{PBHconstraint}, 
%~\cite{Green:2004wb,Frampton:2009nx,Carr:2009jm,Carr:2016hva,Carr:2016drx,Poulter:2019ooo,Wang:2019kaf,Tisserand:2006zx,Graham:2015apa,Koushiappas:2017chw,Authors:2019qbw,DeLuca:2020qqa,Serpico:2020ehh},
which can lead to fruitful phenomena. 
For instance, the detection of GWs by LIGO/Virgo has revealed the ubiquitous existence of BHs 
with $10\sim100M_\odot$, which revives our interest in the hypothesis that such BH binaries 
are formed by
PBHs~\cite{LIGO-PBH}. 
%\cite{Bird:2016dcv,Clesse:2016vqa,Sasaki:2016jop,Chen:2016pud,Blinnikov:2016bxu,Ali-Haimoud:2016mbv,Zumalacarregui:2017qqd,Garcia-Bellido:2017imq}. 
It is also probable that PBHs are the dominant component of the dark matter 
if they are in the mass range $10^{19}\sim10^{22}~\text{g}$, 
which avoids recent observational
 constraints~\cite{PBH-DM-constraint}. 
 %~\cite{Niikura:2017zjd,Katz:2018zrn,Montero-Camacho:2019jte,Sugiyama:2019dgt,Laha:2019ssq,DeRocco:2019fjq}. 
 In addition, PBHs may seed the galaxy or structure formation~\cite{galaxyformation}, the planet 9~\cite{planet9}, or they may provide a mechanism for baryogenesis~\cite{qcd}.
% ~\cite{Bean:2002kx,Kawasaki:2012kn,Nakama:2017xvq,Carr:2018rid,Nakama:2019htb}, the planet 9~\cite{Scholtz:2019csj,Witten:2020ifl}, or they may provide a mechanism for baryogenesis~\cite{Byrnes:2018clq,Carr:2019hud,Garcia-Bellido:2019vlf,Carr:2019kxo,Hajkarim:2019nbx}. 
 For review of PBHs and their observational constraints, see~\cite{Sasaki:2018dmp,Carr:2020gox}.

To generate an amount of PBHs comparable to the energy density of cold dark matter, 
the spectrum of the primordial scalar perturbation is required to have an amplitude of order $10^{-2}$.
 Such a spectrum with a high peak can be realized in various models of  inflation, 
 for instance, models with potential having a near-inflection point~\cite{Yokoyama:1998pt,Cheng:2016qzb,Garcia-Bellido:2017mdw,Cheng:2018yyr,Dalianis:2018frf,Gao:2018pvq,Tada:2019amh,Xu:2019bdp,Mishra:2019pzq,Liu:2020oqe,Atal:2019erb}, 
 modified gravity~\cite{Kannike:2017bxn,Pi:2017gih,Cheong:2019vzl,Cheong:2020rao,Fu:2019ttf,Dalianis:2019vit,Lin:2020goi,Fu:2019vqc}, 
  multi-field inflation~\cite{GarciaBellido:1996qt,Kawasaki:1997ju,Frampton:2010sw,Clesse:2015wea,Inomata:2017okj,Inomata:2017vxo,Palma:2020ejf,Fumagalli:2020adf,Espinosa:2017sgp,Braglia:2020eai,Kawasaki:2019hvt}, 
  curvaton scenarios~\cite{Kawasaki:2012wr,Kohri:2012yw,Ando:2017veq,Ando:2018nge}, 
models with parametric resonance~\cite{Cai:2018tuh,Chen:2020uhe,Chen:2019zza,Cai:2019jah,Cai:2019bmk}, etc..
The induced GWs from these models may be detected
in the near future by the current and future detectors like LIGO/Virgo, KAGRA~\cite{Somiya:2011np}, LISA~\cite{LISA}, 
%\cite{AmaroSeoane:2012km,AmaroSeoane:2012je,Audley:2017drz,Barausse:2020rsu}, 
Taiji~\cite{Guo:2018npi}, Tianqin~\cite{Luo:2015ght}, ET~\cite{ET}, 
%~\cite{Punturo:2010zz,Sathyaprakash:2012jk}, 
DECIGO~\cite{Kawamura:2011zz}, BBO~\cite{BBO}, 
%\cite{Crowder:2005nr,Corbin:2005ny,Baker:2019pnp}, 
FAST~\cite{Lu:2019gsr}, and SKA~\cite{SKA}. 
%\cite{Janssen:2014dka,Maartens:2015mra}. 
The peak frequency of the induced GWs is related to the associated PBH mass as $f_\text{peak}=6.7\times10^{-9}(M_\text{PBH}/M_\odot)^{-1/2}~\text{Hz}$~\cite{Saito:2008jc}. 
If LIGO detections are PBHs, the induced GWs peak at nHz, right in the detection range 
of pulsar timing array (PTA)~\cite{PTAtheory}. 
%~\cite{Sazhin:1977tq,Detweiler:1979wn}.
The fact that there is no detection at present has already started to constrain the PBHs-as-LIGO-BHs
 scenario~\cite{PTAconstraint}. 
 %~\cite{Lentati:2015qwp,Inomata:2016rbd,Orlofsky:2016vbd,Cai:2019elf,Chen:2019xse}. 
For asteroid-mass PBHs, the associated induced GWs peak at $\sim10^{-3}~\text{Hz}$, 
which must be detectable by LISA if PBHs are the dark matter~\cite{Cai:2018dig}.

The spectrum of the induced GWs from a $\delta$-function peak of the scalar perturbation was studied in Refs.~\cite{Ananda:2006af,Saito:2008jc,Alabidi:2012ex,Alabidi:2013wtp},
and an analytical expression for the GW spectrum was derived in Ref.~\cite{Kohri:2018awv}.
However, it was found that the infrared scaling of the GW spectrum depends sensitively on the width, 
which is $f^2$ for a $\delta$-function peak, while $f^3$ for a finite width,
and a transition from $f^3$ to $f^2$ was observed for a finite but small 
width~\cite{Cai:2019cdl,Yuan:2019wwo}. 
In the case of a broad peak, the induced GW spectrum was numerically studied and 
found to have a broad-peak-like feature near its maximum~\cite{Bugaev:2010bb,Byrnes:2018txb,Inomata:2018epa}. 
The goal of this paper is to derive an analytic formula for the induced GW spectrum 
due to the scalar perturbation with the spectrum having a lognormal peak, for both narrow and 
broad widths. The result will provide a very useful tool for studying various phenomena
associated with the original curvature perturbation.

This paper is organized as follows. In Section \ref{sec:setup}, we collect the 
necessary formulae needed to calculate the induced GWs. 
In Section \ref{sec:apply}, we apply the formulae to the curvature perturbation with a lognormal peak. 
In Section \ref{sec:narrow} and \ref{sec:broad}, we consider narrow and broad widths
and present analytic expressions for the GW spectrum, respectively.
We find they agree very well with the numerical results.
We summarize our findings in Section \ref{sec:con}. 
As the analytical calculation is a little bit long and tedious,
readers may directly go to \eqref{Omega4} and  \eqref{Omega10} to check the main results
and Section \ref{sec:con} for the conclusions.

\section{setup}\label{sec:setup}
In linear cosmological perturbation, the perturbed metric can be written in the Newton
gauge as
\be
ds^2=a^2(\eta)\left[-(1+2\Psi)d\eta^2+((1+2\Phi)\delta_{ij}+h_{ij})dx^idx^j\right],
\ee
where $\Psi$ is the Newton potential, $\Phi$ is the intrinsic curvature perturbation, and $\Phi=-\Psi$ for negligibe anisotropic stress~\cite{Bardeen:1980kt},
 $h_{ij}$ is the transverse traceless (tensor) part of the metric perturbation. 
 In momentum space, the tensor perturbation can be written as 
\be
h_{ij}(\eta,\bx)=\int\frac{d^3k}{(2\pi)^{3/2}}
\sum_{\lambda=+,\times}e_{ij}^\lambda(\hat k)h_{\bk,\lambda}(\eta)e^{i\bk\cdot\bx},
\ee
where $e_{ij}^{\lambda}(\hat k)$ are two orthonormal polarization tensors perpendicular to
 $\hat k$-direction. 
The equation of motion for the tensor perturbation of each polarization at second order is sourced by the scalar perturbation:
\be
h_{\bk,\lambda}''+2\mathcal Hh_{\bk,\lambda}'+k^2h_{\bk,\lambda}
=\mathcal S_\lambda(\bk,\eta)=\int d^3\ell~e_{ij}^\lambda(\bk)q^iq^j
f(\bk,\bl,\eta)\Psi_\bl\Psi_{\bk-\bl},
\ee
where the source term, $\mathcal S_\lambda(\bk,\eta)$, is the convolution of the scalar perturbation,
 which can be factorized into the primordial part $\psi_\bl\psi_{\bk-\bl}$ and the combination of the transfer function $T(\ell,\eta)$:
\be
f(\bk,\bl,\eta)=6T(|\bk-\bl|\eta)T(\ell\eta)
+3\eta\frac{\partial T(|\bk-\bl|\eta)}{\partial\eta}T(\ell\eta)
+2\eta^2\frac{\partial T(|\bk-\bl|\eta)}{\partial\eta}\frac{\partial T(\ell\eta)}{\partial\eta}\,;
\quad \Psi(\bk,\eta)=T(k\eta)\psi_\bk\,.
\ee
In this paper we only consider the wavenumbers that reenter the Hubble horizon during
the radiation-dominated stage. For the case of  an (early) matter-dominated stage, see Refs.~\cite{Assadullahi:2009nf,Kohri:2018awv,Inomata:2019ivs,Inomata:2019zqy,Gong:2019mui,Inomata:2020lmk}, while for an arbitrary background, see Refs.~\cite{Bhattacharya:2019bvk,Domenech:2019quo} 
and a companion paper \cite{Domenech:2020kqm}. 
The power spectra of the primordial scalar perturbation and the tensor perturbation are defined as
\begin{align}\label{def:Ppsi}
\langle\psi_\bk\psi_\mathbf{q}\rangle&=\frac{2\pi^2}{k^3}\mathcal{P}_\psi(k)\delta^{(3)}(\bk+\bp).\\
\label{def:Ph}
\sum_{\lambda=+,\times}\langle e_{ij}^\lambda e^{ij}_\lambda h_{\bk,\lambda}(\eta)h_{\bp,\lambda}(\eta)\rangle&=\frac{2\pi^2}{k^3}\mathcal{P}_h(k,\eta)\delta^{(3)}(\bk+\bp).
\end{align}
The power spectrum of the primordial Newton potential is related to that of the conserved comoving 
curvature perturbation in the radiation dominated universe as $\mathcal{P}_\psi=(4/9)\mathcal{P}_\mathcal{R}$.
As mentioned in Introduction, the constraint by Planck is 
$\mathcal{P}_\mathcal{R}(k)\approx2.1\times10^{-9}$ at $k=0.05~\text{Mpc}^{-1}$~\cite{Aghanim:2018eyx}. 
However, since the scales we have in mind are substantially smaller than 1 Mpc,
we are free from the Planck constraint. 
We will specify the form of $\mathcal{P}_\mathcal{R}(k)$ on small scales later. 

The equation of motion for $h_\bk$ can be solved by the Green function method. 
In the radiation-dominated universe, %~\cite{Baumann:2007zm}:
\be\label{sol:h}
h_{\bk,\lambda}(\eta)=\int^\eta_{\eta_0} d\tilde\eta
\frac{\sin k(\eta-\tilde\eta)}{k}\frac{a(\tilde\eta)}{a(\eta)}\mathcal S(\bk,\tilde\eta).
\ee
The two-point correlation function of the tensor perturbation is given by
\begin{align}\nn
&\langle h_{\bk,\lambda}(\eta)h_{\bp,\lambda'}(\eta)\rangle=
\delta_{\lambda,\lambda'}
\int^\eta_{\eta_0}d\eta_2\int^\eta_{\eta_0}d\eta_1\frac{a(\eta_2)}{a(\eta)}
\frac{a(\eta_1)}{a(\eta)}\frac{\sin k(\eta-\eta_1)}{k}\frac{\sin p(\eta-\eta_2)}{p}
f(\bk,\bl,\eta_1)f(\bp,\bq,\eta_2)
\\\label{def:<hh>}
&\qquad\times\left(\frac49\right)^28\pi^4\int\frac{d^3\ell}{(2\pi)^{3/2}}e_{ij}^\lambda(\hat k)\ell^i\ell^j
\int\frac{d^3q}{(2\pi)^{3/2}}e_{\ell m}^{\lambda'}(\hat p)q^\ell q^m
\frac{\mathcal{P}_\mathcal{R}(l)\mathcal{P}_\mathcal{R}(|\bk-\bl|)}{\ell^3|\bk-\bl|^3}
\delta^{(3)}\left(\bk+\bp\right)
\delta^{(3)}\left(\bl+\bq\right).
\end{align}
Here it may be worth mentioning that we do not assume anything about the statistical nature
of the primordial scalar perturbation. It may be Gaussian or may be highly non-Gaussian.
The essential point is that the spectrum of the induced GWs is given by a convolution
of the primordial curvature perturbation spectrum, independent of its statistical nature.
Nevertheless, we also mention that it may be possible to obtain the information about
the non-Gaussianity if we combine the predictions on the induced GWs and those on
the corresponding abundance of PBHs~\cite{Cai:2018dig}.

Cosmologists commonly call the GW energy density per logarithmic interval of wavenumber, 
normalized by the total energy density of the universe as the GW spectrum, 
\be
\Omega_\text{GW}(k,\eta)=\frac1{\rho_\text{tot}}\frac{d\rho_\text{GW}}{d\ln k}=\frac{k^2}{12H^2a^2}\mathcal{P}_h(k,\eta).
\ee
The power spectrum of the tensor perturbation is defined in \eqref{def:Ph} by the 2-point correlation function, 
which is expressed in \eqref{def:<hh>}. 
To step further we have to calculate the time integral of the transfer functions, and average it over many oscillating periods, which is done in Refs.~\cite{Espinosa:2018eve,Kohri:2018awv,Gong:2019mui}. 
Following Ref.~\cite{Kohri:2018awv}, we define a set of new variables, $u=|\bk-\bl|/k$, and $v=\ell/k$, 
and then integrate the solid angular elements perpendicular to $\hat{\ell}$- and $\hat{q}$-direction in \eqref{def:<hh>},
\begin{align}\label{Omega0}
\Omega_\text{GW,r}(k)
&=3\int^\infty_0dv\int^{1+v}_{|1-v|}du\frac{\mathcal{T}(u,v)}{u^2v^2}\mathcal{P}_\mathcal{R}(vk)\mathcal{P}_\mathcal{R}(uk),\\\nn
\mathcal{T}(u,v)&=\frac14\left[\frac{4v^2-(1+v^2-u^2)^2}{4uv}\right]^2\left(\frac{u^2+v^2-3}{2uv}\right)^4
\left[\left(\ln\left|\frac{3-(u+v)^2}{3-(u-v)^2}\right|-\frac{4uv}{u^2+v^2-3}\right)^2+\pi^2\Theta\left(u+v-\sqrt3\right) \right],
%T(u,v)&=\frac14\left(\frac{4v^2-(1+v^2-u^2)^2}{4uv}\right)^2\left(\frac{u^2+v^2-3}{2uv}\right)^2\\
%&\cdot\left\{\left[-2+\frac{u^2+v^2-3}{2uv}\ln\left|\frac{3-(u+v)^2}{3-(u-v)^2}\right|\right]^2+\pi^2\left(\frac{u^2+v^2-3}{2uv}\right)^2\Theta\left(u+v-\sqrt3\right) \right\}.
\end{align}
$\Theta$ is the step function. The normalization of $\mathcal{T}(u,v)$ is such that when $u\approx v\rightarrow\infty$, we have $\mathcal{T}(u,v)\rightarrow(\ln(u+v))^2/4\sim(\ln v)^2$. $\Omega_\text{GW,r}(k)$ in \eqref{Omega0} is the spectrum during the radiation dominated era which does not depend on time as it is the asymptotic amplitude averaged over many periods deep inside the horizon. The subscript ``r'' is added to avoid possible confusion with time-dependent $\Omega_\text{GW}(k,\eta)$. This \textit{short wavelength limit} is crucial in the recent discussions on the gauge dependence of the induced GWs~\cite{Inomata:2019yww,othergaugepaper}. 
For any form of the power spectrum of the curvature perturbation, the spectrum of the induced GWs in the radiation-dominated stage can be calculated by numerically integrating \eqref{Omega0}. 
However, we note that since the GW energy density starts to decay relative to the matter density
after the matter-radiation equality, the observational GW spectrum today is given by 
\begin{align}\label{Omega6}
\Omega_\text{GW}(f,\eta_0)h^2
=\frac{g_*(\eta_0)^{4/3}}{g_*(\eta_0)g_{*s}(\eta_k)^{1/3}}\Omega_{r,0}\Omega_\text{GW,r}(f)
=1.6\times10^{-5}\left(\frac{g_{*s}(\eta_k)}{106.75}\right)^{-1/3}\left(\frac{\Omega_{r,0}h^2}{4.1\times10^{-5}}\right)\Omega_\text{GW,r}(f).
\end{align}
Here $\Omega_\text{GW,r}(f)$ is the GW spectrum at matter-radiation equality given by \eqref{Omega0}, and 
to compare with the observables we replaced the comoving wavenumber $k$ to 
the physical frequency $f$, by $f=k/(2\pi a_0)$. Explicitly, 
$f=1.5\times 10^{-9}(k/1\,{\rm pc}^{-1}){\rm Hz}$. 

 In the following sections, we calculate $\Omega_\text{GW,r}$ in \eqref{Omega0} 
 for both narrow and broad widths of a lognormal peak in the power spectrum of the 
 primordial curvature perturbation. 
 Comparison of the result with observation signals should be done by using \eqref{Omega6}
 for the current GW spectrum $\Omega_\text{GW}(f,\eta_0)$. 
 %From now on, we will omit the superscipt $r$ of $\Omega_{\text{GW}}^r$ for clearance.
 %throughout the paper, while \eqref{Omega6} should be keep in mind when we want to apply our result to 

\section{Lognormal Peak}\label{sec:apply}
In the literature, a lognormal peak in the power spectrum of the curvature perturbation is often considered, %for the PBH formation, 
\be\label{lognormalpeak}
\mathcal{P}_\mathcal{R}(k)=\frac{\mathcal{A_R}}{\sqrt{2\pi}\Delta}
\exp\left(-\frac{\ln^2(k/k_*)}{2\Delta^2}\right),
\ee
where $k_*$ is the peak wavenumber, $\Delta$ is the dimensionless width, and the coefficient is to satisfy 
the normalization $\int^\infty_0\mathcal{P}_\mathcal{R}(k)\,d\ln k=\mathcal{A_R}$. A lognormal peak in the power spectrum of the primordial curvature perturbation can arise naturally in many inflation models (see below), and it is always a good fit to a peaked spectrum around its peak. Furthermore,  
the scale-invariant power spectrum responsible for the CMB anisotropies can be included as the infinitely 
broad limit: $\mathcal{A}_\mathcal{R},\Delta\rightarrow\infty$ while keeping $\mathcal{A}_\mathcal{R}/\Delta=\text{constant}$. Another limit is $\Delta\rightarrow0$, when \eqref{lognormalpeak} reduces to the $\delta$-function peak, i.e. $\mathcal{P}_\mathcal{R}=\mathcal{A}_\mathcal{R}\delta(\ln(k/k_*))$. 
 This is used in some papers of generating PBHs which we will comment more in Sec.\ref{sec:narrow}. 
For the lognormal peak, \eqref{Omega0} reduces to 
\be\label{Omega1}
\Omega_\text{GW,r}=\frac{3}{2\pi}\frac{\mathcal A_\mathcal{R}^2}{\Delta^2}
\int^\infty_0dv\int^{1+v}_{|1-v|}du\frac{\mathcal{T}(u,v)}{u^2v^2}\exp\left[-\frac{(\ln u)^2+(\ln v)^2+2\ln\kappa\ln(uv)+2\ln^2\kappa}{2\Delta^2}\right],
\ee
where we have defined $\kappa\equiv k/k_*$ for convenience. 
This GW spectrum is valid until the moment of matter-radiation equality, 
which is connected to the GW spectrum today by \eqref{Omega6}.

\begin{figure}[htbp]
\begin{center}
\includegraphics[width=0.6\textwidth]{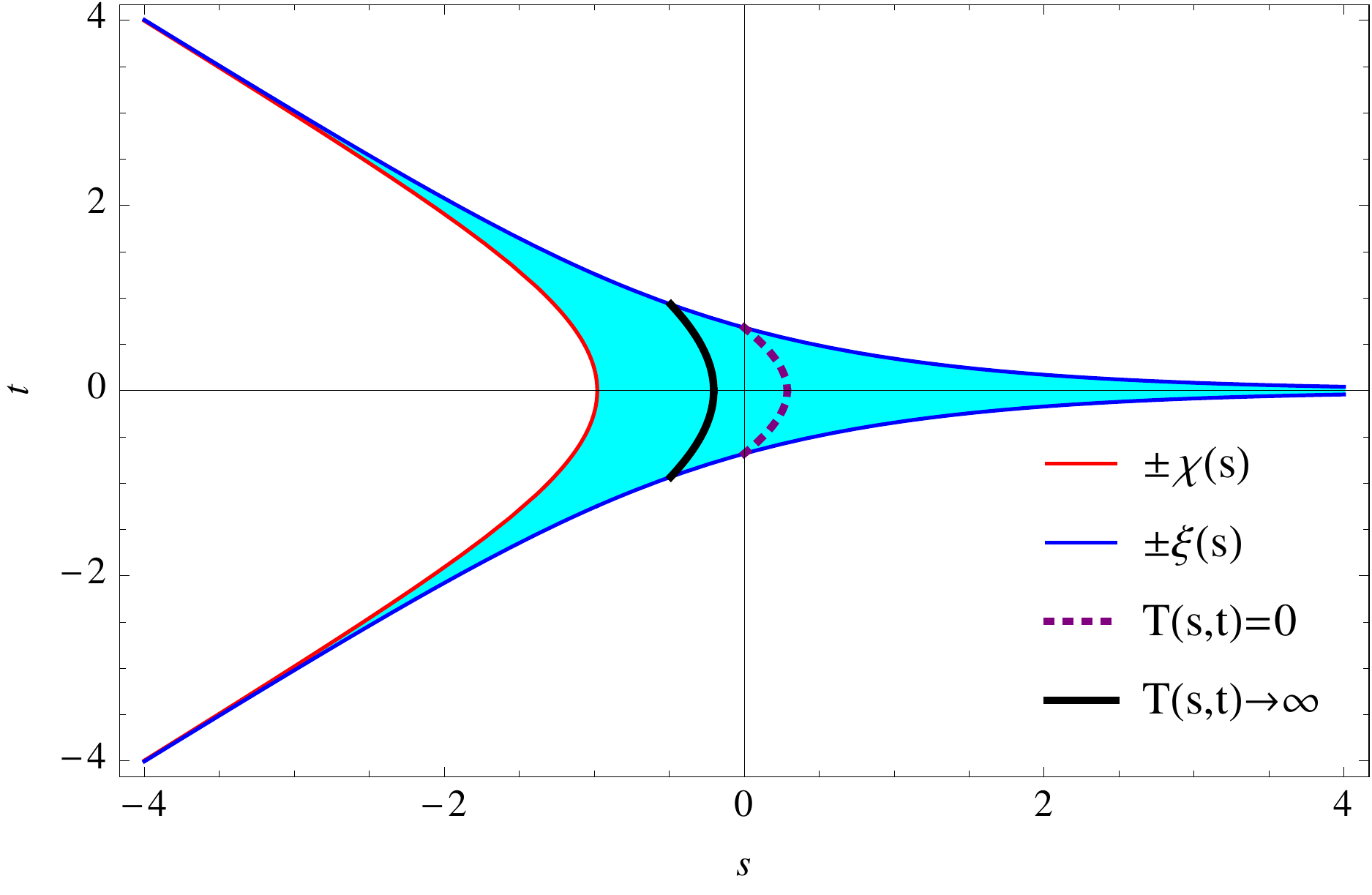}
\caption{The shaded area is the domain of integration \eqref{Omega3}, 
	which is bounded by the curves $\pm\xi(s)$ and $\pm\chi(s)$ shown in \eqref{boundary}. 
	We also show the curves of zero points (dotted purple) and 
	the logarithmic divergence (black) in the integral kernel $\mathscr{T}(s,t)$ defined in \eqref{def:Tst}. 
	The former curve is given by $e^{\sqrt2s}\cosh(\sqrt2t)=3/2$, while the latter curve is $e^{\sqrt2s}\cosh^2(t/\sqrt2)=3/4$. }\label{fig:area}
\label{default}
\end{center}
\end{figure}

It is convenient to define the following new varables:
\be
s=\frac1{\sqrt2}\ln(uv),\qquad
t=\frac1{\sqrt2}\ln\frac uv.
%\;\;\;\;\left(
%\text{Inversely}~
%\alpha=\frac{s+t}{\sqrt2},\;\;\;\;
%\beta=\frac{s-t}{\sqrt2}.\right)
\ee
Under this coordinate transformation, \eqref{Omega1} becomes
%Then we can simplify write the integral \eqref{Omega2} as
\begin{align}\label{Omega3}
\Omega_\text{GW,r}&=\frac{3}{\pi}\frac{\mathcal A_\mathcal{R}^2}{\Delta^2}\kappa^2e^{\Delta^2}
\int^\infty_{-\infty}\exp\left[-\frac{\left(s+\sqrt2(\ln\kappa+\Delta^2)\right)^2}{2\Delta^2}\right]ds
\int^{\xi(s)}_{\chi(s)}\mathscr{T}\left(s,t\right)\exp\left(-\frac{t^2}{2\Delta^2}\right)dt,\\\nn
%&\quad\quad\quad\quad\cdot\left(\int^{\xi(s)}_{-\xi(s)}\mathscr{T}\left(s,t\right)\exp\left(-\frac{t^2}{2\Delta^2}\right)dt-\int^{\chi(x)}_{-\chi(s)}\mathscr{T}\left(s,t\right)\exp\left(-\frac{t^2}{2\Delta^2}\right)dt\right),\\\nn
\mathscr{T}(s,t)
&=\frac14\left(\cosh(\sqrt2t)-\frac14e^{-\sqrt2s}-e^{\sqrt2s}\sinh^2(\sqrt2t)\right)^2
\left(\cosh(\sqrt2t)-\frac32e^{-\sqrt2s}\right)^4\\\label{def:Tst}
&\times\left\{\left[\ln\left|\frac{3-4e^{\sqrt2s}\cosh^2\left(t/\sqrt2\right)}{3-4e^{\sqrt2s}\sinh^2\left(t/\sqrt2\right)}\right|
-\frac2{\cosh(\sqrt2t)-\frac32e^{-\sqrt2s}}
\right]^2+\pi^2\Theta\left(2e^{\frac{s}{\sqrt2}}\cosh\frac{t}{\sqrt2}-\sqrt3\right) \right\}.
\end{align}
The integral on $t$ is bounded by the lines $v=1+u$ and $v=|1-u|$, which correspond to
 the following curves in the new coordinates,
\be\label{boundary}
\chi(s)=\mathbf{Re}\left[\sqrt{2} \text{arccosh}\left(\frac{e^{-s/\sqrt2}}{2}\right)\right],
\quad
\xi(s)=\sqrt{2} \text{arcsinh}\left(\frac{e^{-s/\sqrt2}}{2}\right),
\ee
where the real part is taken to ensure that $\chi(s)=0$ for $s>-\sqrt2\ln2$. 
The domain is shown in Fig.~\ref{fig:area}. 
As the integrand is proportional to Gaussian functions, the result 
depends crucially on whether the widths of the Gaussian peaks are inside the integration
 domain or not, which are determined by  the values of $\Delta$.
  We will discuss the cases of $\Delta\ll1$ and $\Delta\gtrsim1$ separately in the following subsections.

\subsection{Narrow Peak ($\Delta\ll1$)}\label{sec:narrow}
\begin{figure}[htbp]
\begin{center}
\includegraphics[width=0.48\textwidth]{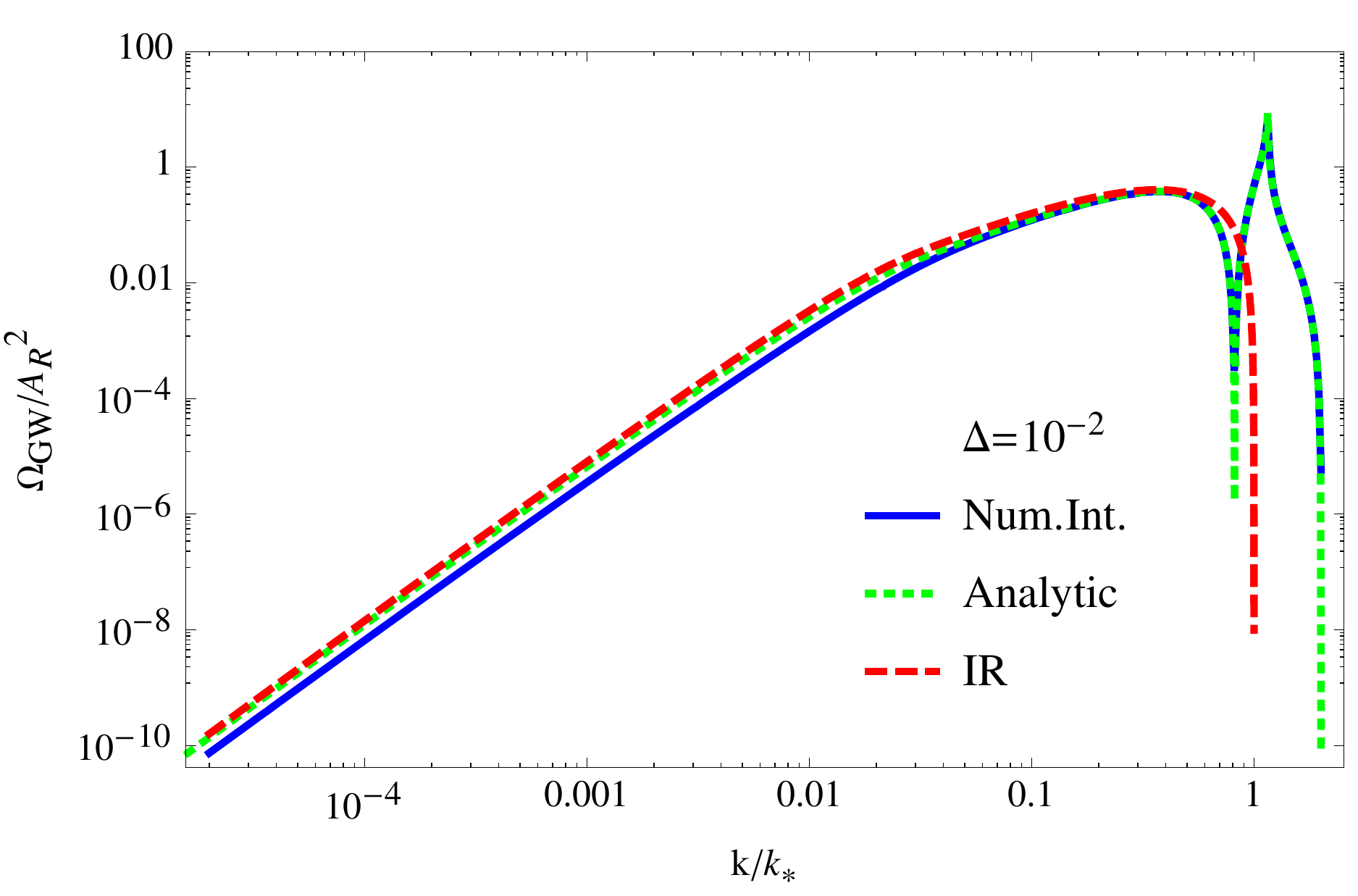}
\includegraphics[width=0.48\textwidth]{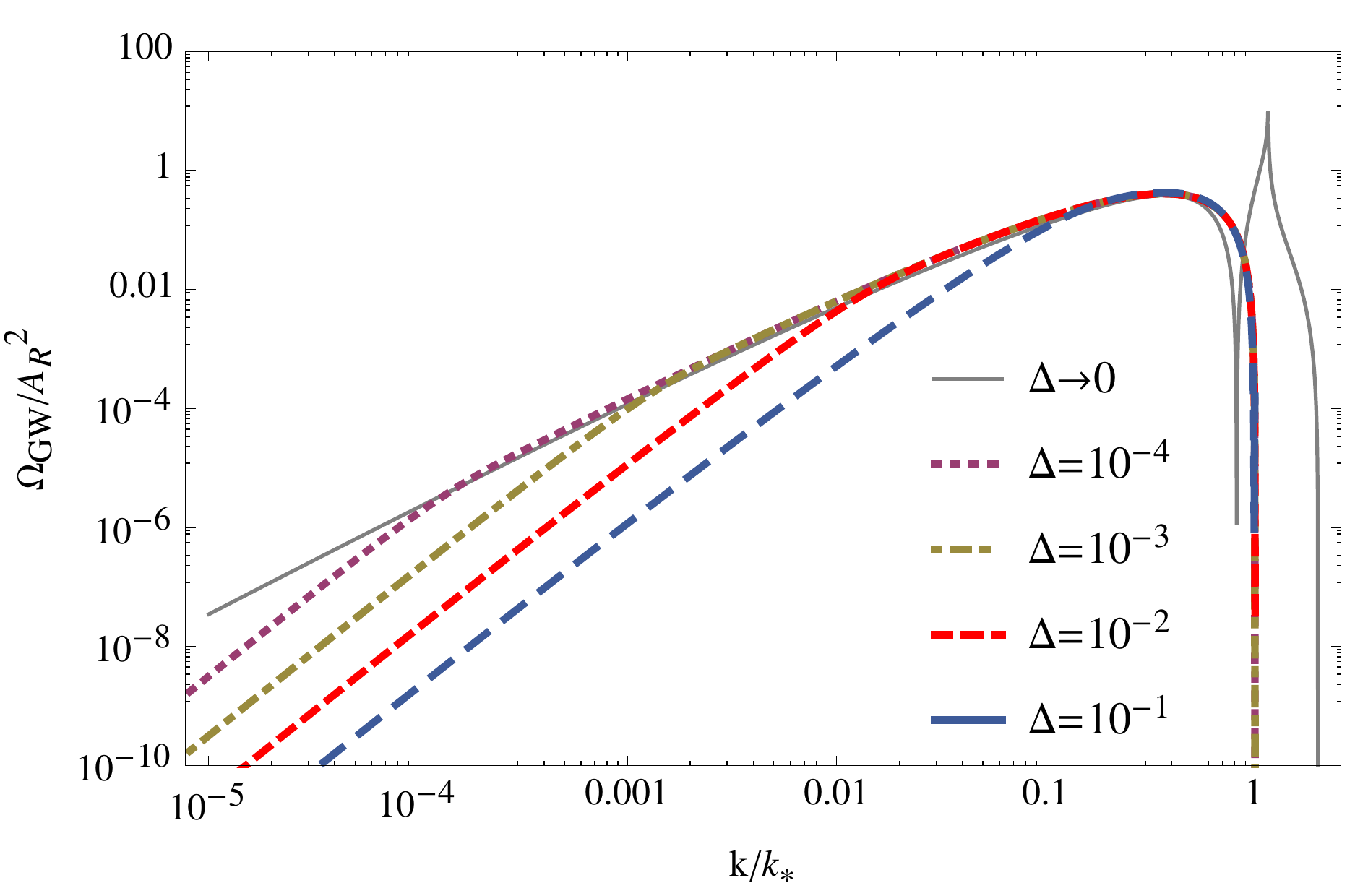}
\caption{Left: The GW spectrum with $\Delta=10^{-2}$. 
The results of numerical integration of \eqref{Omega0}, 
the analytical expression \eqref{Omega4}, and the infrared approximation \eqref{Omega5}
are shown.
 Right: The GW spectra for different values of widths according to the infrared approximation \eqref{Omega5}. 
The dashed gray curve is the GW spectrum from a $\delta$-function peak drawn for reference, whose near-peak behavior is nearly the same for all the cases  of $\Delta\ll1$. } \label{fig:narrow}
\end{center}
\end{figure}

Some models predict a narrow peak in the curvature perturbation,
$\Delta\ll1$~\cite{Kawasaki:1997ju,Frampton:2010sw,Kawasaki:2012wr,Pi:2017gih,Inomata:2017okj,Cai:2018tuh,Chen:2020uhe,Chen:2019zza,Cai:2019jah}. 
In this case, the main contribution of the integral \eqref{Omega3} on $s$ and $t$ comes from the vicinity of the peak, $s=-\sqrt2(\ln\kappa+\Delta^2)$ and $t=0$. 
For the integral on $t$ in the second line of \eqref{Omega3}, 
we can take $t\rightarrow0$ in $\mathscr{T}(s,t)$, and then perform the Gaussian integral,
\be\label{tintegralnarrow}
2\int^{\xi(s)}_{\chi(s)}\mathscr{T}(s,t)\exp\left(-\frac{t^2}{2\Delta^2}\right)dt
\approx\mathscr{T}(s,0)\sqrt{2\pi}\Delta\left[\text{erf}\left(\frac{\xi(s)}{\sqrt2\Delta}\right)-\text{erf}\left(\frac{\chi(s)}{\sqrt2\Delta}\right)\right].
\ee
We substitute this back into \eqref{Omega3}. Again the narrow Gaussian distribution guarantees that we can take $s\rightarrow-\sqrt2(\ln\kappa+\Delta^2)$ in \eqref{tintegralnarrow}, and then perform the Gaussian integral on $s$. We then obtain an analytical result for the GW spectrum in the narrow-peak case,
\begin{align}\nn
\Omega_\text{GW,r}&\approx3\mathcal{A}_\mathcal{R}^2\kappa^2e^{\Delta^2}
\left[\text{erf}\left(\frac{1}{\Delta}\text{arcsinh}\frac{\kappa e^{\Delta^2}}2\right)-
\text{erf}\left(\frac{1}{\Delta}\mathbf{Re}\left(\text{arccosh}\frac{\kappa e^{\Delta^2}}2\right)\right)\right]\left(1-\frac14\kappa^2e^{2\Delta^2}\right)^2
\left(1-\frac32\kappa^2e^{2\Delta^2}\right)^2\\\label{Omega4}
&\quad\cdot\left\{\left[\frac12\left(1-\frac32\kappa^2e^{2\Delta^2}\right)\ln\left|1-\frac4{3\kappa^2e^{2\Delta^2}}\right|-1\right]^2
+\frac{\pi^2}4\left(1-\frac32\kappa^2e^{2\Delta^2}\right)^2\Theta\left(2-\sqrt3\kappa e^{\Delta^2}\right)
\right\},
\end{align}
where $\kappa\equiv k/k_*$ as defined before.
This analytical result fits quite well with the numerical integration of \eqref{Omega1}, 
which is shown in Fig.~\ref{fig:narrow} for $\Delta=10^{-2}$. 
In the limit $\Delta\rightarrow0$, the result reduces to the GW spectrum induced by
a $\delta$-function peak, obtained in Ref.~\cite{Kohri:2018awv},
\begin{align}\label{Omega2}
\Omega_\text{GW,r}^{(\delta)}&=3\mathcal{A}_\mathcal{R}^2\kappa^2
\left(1-\frac14\kappa^2\right)^2\left(1-\frac32\kappa^2\right)^4
\left[\left(\frac12\ln\left|1-\frac4{3\kappa^2}\right|-\frac1{1-\frac32\kappa^2}\right)^2
+\frac{\pi^2}4\Theta\left(2-\sqrt3\kappa\right)\right]\Theta(2-\kappa).
\end{align}

Our narrow-peak result \eqref{Omega4} can be further simplified if we approximate 
$e^{\Delta^2}\approx1$,
\be\label{main1}
\Omega_{\text{GW,r}}^{(\Delta\ll1)}\approx\text{erf}
\left(\frac1\Delta\text{arcsinh}\frac{k}{2k_*}\right)\Omega_{\text{GW,r}}^{(\delta)}.
\ee
We note that the error function factor comes from \eqref{tintegralnarrow} which is independent of
 the transfer function \eqref{main1}, implying that this formula is valid for any universe
 with arbitrary equation of state $w$. The whole equation-of-state dependence is contained in
 $\Omega_{\text{GW,r}}^{(\delta)}$.
 In a companion paper \cite{Domenech:2020kqm}, an intermediate stage
 dominated by a scalar field with constant $w$ between inflation and 
 the radiation-dominated universe is considered, 
 where \eqref{main1} is applied to find the infrared behavior of the induced GWs from a narrow peak.

To summarize, the infrared behavior of GW spectrum is given by \eqref{Omega4} or \eqref{main1} 
by setting $\kappa\ll1$,
\begin{align}\label{Omega5}
\frac{\Omega_\text{GW,r}^{\text{(IR)}}}{\mathcal{A}_\mathcal{R}^2}
\approx3\kappa^2e^{\Delta^2}
%\left[1+6\Delta^2\left(1-\exp\left(-\frac{\kappa^2e^{2\Delta^2}}{4\Delta^2}\right)\right)\right]
\Big(\ln\kappa+\Delta^2\Big)^2\text{erf}\left(\frac{\kappa e^{\Delta^2}}{2\Delta}\right)
\approx\left\{
\begin{matrix}
\displaystyle 3\kappa^3\frac{e^{2\Delta^2}}{\sqrt\pi\Delta}\left(\ln\kappa+\Delta^2\right)^2, & \text{for}~\kappa< 2\Delta e^{-\Delta^2}~\text{(far-IR)};\\
&\\
3\kappa^2e^{\Delta^2}\left(\ln\kappa+\Delta^2\right)^2, & \text{for}~2\Delta e^{-\Delta^2}<\kappa<1~\text{(near-IR)}.
\end{matrix}
\right.
\end{align}
In the last step, for an intuitive understanding of the scaling law of the GW spectrum, we use the asymptotic expression for the error function for small and large arguments, respectively, 
which we refer to as the far-IR and near-IR regions. 
As clear from the above as well as from Fig.~\ref{fig:narrow}, 
there is a break point at $k/k_*=2\Delta e^{-\Delta^2}$ where the GW spectrum changes 
its infrared behavior from $\Omega_\text{GW,r}\propto k^3$ to $\Omega_\text{GW,r}\propto k^2$. 
Let us denote this break wavenumber by $k_b$, i.e., $k_b=2\Delta e^{-\Delta^2}k_*$.
This feature is characteristic of the GW spectrum induced by a narrowly peaked scalar perturbation, 
and it is a very useful feature in both identifying the origin of the curvature perturbation 
and measuring the peak width.

It is clear to see in \eqref{Omega4} that besides an infrared local maximum at around $k/k_*=1/e$, there is another ``resonance peak'' with logarithmic divergence,
 which is located at $k=(2/\sqrt3)e^{-\Delta^2}k_*$, which we denote by $k_p$. 
 The two maxima are separated by a dip at $k/k_*=\sqrt{2/3}e^{-\Delta^2}$. 
 This two-peak structure is another characteristic feature of the induced GW from a narrow peak, 
 which can be obviously seen for $\Delta\lesssim0.4$.
\footnote[4]{The origin of the dip and the sharp peak at $k=k_p$
	is due to  resonance of the tensor modes and scalar modes inside the Hubble horizon
	due to the difference in their sound speeds~\cite{Ananda:2006af}. 
	If the sound speeds for both tensor modes and scalar modes are the same, 
	the resonance is absent, and we only have a single smooth peak in the GW spectrum.
	See Ref.~\cite{Domenech:2020kqm}.} 

The value at the infrared local maximum can be easily estimated as
\be\label{IRmax}
\Omega_\text{GW,r}^\text{(IR,max)}\approx\Omega_\text{GW}\left(\frac1e\right)=\frac3{e^2}\mathcal{A}_\mathcal{R}^2e^{\Delta^2}\left(1-e^2\Delta^2\right)^2\approx0.41\mathcal{A}_\mathcal{R}^2,
\ee
where the dependence on the width is of order $\mathcal{O}(\Delta^2)$ and can be neglected. 
Thus the infrared maximum of $\Omega_\text{GW,r}$ only depends on the normalization of the 
curvature perturbation spectrum, which can be clearly seen in the right panel of Fig.\ref{fig:narrow}. 

As for the resonance peak, it must be smoothed when compared to observation
as an infinitely narrow band observation is impossible~\cite{Thrane:2013oya}. 
For instance, Ref.~\cite{Byrnes:2018txb} smooths the resonance peak induced by a $\delta$-function 
peak on one $e$-fold. 
Here we smooth it by taking account of the following fact.
For a finite width peak, its width provides a natural smoothing scale,
\begin{align}
\Omega_\text{GW,r}^\text{(res,max)}
&=\frac1{2\Delta}\int^{k_*e^\Delta}_{k_*e^{-\Delta}}\Omega_\text{GW,r}\frac{dk}{k}
%&\approx\frac49\mathcal{A}_\mathcal{R}^2\left[\left(\ln\Delta\right)^2+2\left(1+\ln2\right)\ln\Delta+\left(1+\ln2\right)^2+1+\frac{\pi^2}{2}\right],\\
\approx\frac49\mathcal{A}_\mathcal{R}^2\left[\Big(\ln(2\Delta)+1\Big)^2+1+\frac{\pi^2}{2}\right].
\end{align}
We can see that if $\Delta$ is not too small, i.e. for $\Delta\gg(1/2)\exp(-\sqrt{4+2\pi^2}-1)\approx0.016$, the smoothed peak value is independent of its width and can be estimated by
\begin{align}
 \Omega_\text{GW,r}^\text{(res,max)}\approx
 \frac{4}{9}\left(1+\frac{\pi^2}{2}\right)
 \mathcal{A}_\mathcal{R}^2\approx2.6\mathcal{A}_\mathcal{R}^2\,.
\end{align}
Note that this is always higher than the infrared peak given in \eqref{IRmax}. 
For $\Delta\lesssim0.016$, the peak is proportional to $(\ln\Delta)^2$, i.e. $\Omega_\text{GW,r}^\text{(res,peak)}\approx(4/9)(\ln\Delta)^2\mathcal{A}_\mathcal{R}^2$. 
If $f_*\Delta$ is smaller than the frequency resolution, $\delta f$, the latter should be
 used instead as the smoothing scales, though it is unlikely that the observation with such high resolution
 becomes possible in the near future. 

\subsection{Broad Peak ($\Delta\gtrsim\mathcal{O}(1)$)}\label{sec:broad}
Many models predict a broad peak in the scalar perturbation, $\Delta\gtrsim\mathcal{O}(1)$~\cite{GarciaBellido:1996qt,Inomata:2017vxo,Inomata:2017okj,Garcia-Bellido:2017mdw,Kannike:2017bxn,Yokoyama:1998pt,Kohri:2012yw,Clesse:2015wea,Cheng:2016qzb,Ando:2018nge,Cheng:2018yyr,Ando:2017veq,Espinosa:2017sgp,Espinosa:2018eve,Braglia:2020eai}. 
In this case, we show that the induced GW spectrum has a lognormal peak with width $\Delta/\sqrt2$. %, together with %the sharp cutoff at $k<2k_*$ in the narrow-peak case replaced by an exponential suppresion. 
%\textbf{(SP: A lognormal peak with width of $\Delta/\sqrt2$ already has an exponential suppression. However, as you can see from \eqref{Omega11}, the UV suppression goes as a lognormal peak of width $\Delta$. The reason is subtle. I can not think of an easy explanation. To reduce confusion, I suggest to delete this sentence of ``exponential suppresion''.)}
Since the integrand in (\ref{Omega3}) is not concentrated around the vicinity of the peak, 
it is difficult to calculate it at once. 
It is found that $\mathscr{T}(s,t)$ in the integrand behaves differently for 
$s\gtrsim1$, $|s|\sim\mathcal{O}(1)$, and $s\lesssim-1$. 
We therefore decompose the integral into these three different domains, evaluate each
separately, and add up all the contributions at the end to obtain
 a formula that has sufficiently satisfactory accuracy.

First we consider the domain $s\gtrsim1$.
The contribution in the infrared is mainly from large $s$, 
where we may take the leading order of $\mathscr T(s,t)$ for $s\gg(1/\sqrt2)\ln(3/2)$ as
\be
\mathscr{T}(s,t)\approx\frac{s^2}{2}
\left[\cosh\left(\sqrt2t\right)-e^{\sqrt2s}\sinh^2\left(\sqrt2t\right)\right]^2\cosh^4\left(\sqrt2t\right).
\ee
Then since $\xi(s)\ll\Delta$ for $\Delta\gtrsim1$ and $s\gg1$,
the integral along the $t$-axis can be approximated by an error function as
\be\label{tgauss}
\int^\xi_{-\xi}\mathscr{T}(s,t)\exp\left(-\frac{t^2}{2\Delta^2}\right)dt
\approx \frac{8}{15\sqrt2}s^2e^{-s/\sqrt2}.
\ee
Substituting this back into \eqref{Omega3}, and perform the Gaussian integral along the $s$-axis,
we find
\begin{align}\nn
\Omega_\text{GW,r}^{(s\gtrsim1)}
&\approx\frac4{5\sqrt2\pi}\kappa^3\frac{e^{9\Delta^2/4}}{\Delta^2}
\int^\infty_{\frac1{\sqrt2}\ln\frac32}s^2\exp\left[-\frac{\left(s+\sqrt2(\ln\kappa+\frac32\Delta^2)\right)^2}{2\Delta^2}\right]ds,\\\nn
&=\frac{4}{5\sqrt\pi}\mathcal{A}_\mathcal{R}^2\kappa ^3\frac{e^{\frac{9 \Delta ^2}{4}}}{\Delta}
\left\{
\left[\left(\ln\kappa+\frac32\Delta^2\right)^2+\frac{\Delta^2}{2}\right] 
\text{erfc}\left(\frac{\ln\kappa+\frac32\Delta^2+\frac12\ln\frac32}{\Delta}\right)\right.\\\label{Omega7}
&\qquad\qquad
\left.-\frac{\Delta}{\sqrt\pi}\exp\left(-\frac{\left(\ln\kappa+\frac32\Delta^2+\frac12\ln\frac32\right)^2}{\Delta^2}\right)
\left(\ln\kappa+\frac32\Delta^2-\frac12\ln\frac32\right)\right\}.
\end{align}
where $\text{erfc}(x)$ is the complementary error function defined as $\text{erfc}(x)\equiv1-\text{erf}(x)$. 
For infrared $k$, $\kappa\ll1$, the complementary error function approaches 2, 
while the last term in the curly brackets in \eqref{Omega7} is exponentially suppressed.
Thus the infrared GW spectrum is given approximately by
\be\label{IR}
\Omega_\text{GW,r}^\text{(IR)}\approx
\frac8{5\sqrt\pi}\mathcal{A}_\mathcal{R}^2\frac{e^{\frac{9 \Delta ^2}{4}}}{\Delta}
\kappa ^3\left[\left(\ln\kappa+\frac32\Delta^2\right)^2+\frac{\Delta^2}{2}\right]
\qquad\text{for}~\kappa\lesssim\sqrt\frac23\exp\left(-\frac32\Delta^2\right).
\ee
We see the familiar $k^3$ scaling. 
On the other hand, for $\kappa\gtrsim\sqrt\frac23\exp(-\frac32\Delta^2)$, 
the complimentary error function is also exponentially suppressed.
Taking the large $\Delta$ limit of it gives the near-peak behavior,
\begin{align}\label{peak1}
\Omega_\text{GW,r}^{(\text{peak},s\gtrsim1)}
&\approx
\frac{16+24\ln\frac32+18\ln^2\frac32}{135\pi}
\mathcal{A}_\mathcal{R}^2\frac{e^{9\Delta^2/4}}{\Delta^2}
\exp\left(-\frac{\left(\ln\kappa+\frac32\Delta^2+\frac12\ln\frac32\right)^2}{\Delta^2}\right).
%\\\label{peak1}
%&\approx\mathcal{A}_\mathcal{R}^2\frac{0.0368}{\Delta^2}\exp\left(-\frac{\ln^2\kappa}{\Delta^2}\right).
\end{align}

Second, we focus on the domain $|s|\lesssim1$.
There is a discontinuity at 
 $t=\pm\sqrt2\text{arccosh}\left(\frac{\sqrt3}{2}e^{-s/\sqrt2}\right)$ when $|s|\lesssim1$.
 The main contribution is from the right hand side of the singularity, as there is an extra $\pi^2$ term 
 in the curly brackets  in \eqref{def:Tst},  which allows us to take $t\rightarrow0$ and neglect the 
 logarithm term, giving
\be\label{temp20}
\mathscr{T}(s,t)\approx\frac{\pi^2}{4}\left(1-\frac{3}{2} e^{-\sqrt{2} s}\right)^4 \left(1-\frac{1}{4} e^{-\sqrt{2} s}\right)^2.
\ee
As this is independent of $t$, the remaining integral on $t$ is just an error function, which gives
\begin{align}
\Omega_\text{GW,r}^{|s|<1}&\approx\frac3{\pi}\frac{\mathcal{A}_\mathcal{R}^2}{\Delta^2}\kappa^2 e^{\Delta^2}
\int^{\frac1{\sqrt2}\ln\frac32}_{-\frac1{\sqrt2}\ln\frac43}ds
\exp\left[-\frac{\left(s+\sqrt2(\ln\kappa+\Delta^2)\right)^2}{2\Delta^2}\right]
\nonumber\\
&\qquad\qquad\qquad\times\frac{\pi^2}{4} \left(1-\frac{3}{2} e^{-\sqrt{2} s}\right)^4 
\left(1-\frac{1}{4} e^{-\sqrt{2} s}\right)^2
\sqrt{\frac\pi2}\Delta\text{Erf}\left(\frac{\xi(s)}{\sqrt2\Delta}\right),
\nonumber\\
&\approx\frac{3\pi}{2\sqrt2}\frac{\mathcal{A}_\mathcal{R}^2}{\Delta^2}\kappa^2 e^{\Delta^2}\exp\left[-\frac{\left(\ln\kappa+\Delta^2\right)^2}{\Delta^2}\right]
\int^{\frac1{\sqrt2}\ln\frac32}_{-\frac1{\sqrt2}\ln\frac43}ds~
e^{-2 \sqrt{2} s} \left(1-2 e^{\sqrt{2} s}\right)^2\text{arcsinh}\left(\frac{e^{-s/\sqrt2}}{2}\right).
\label{small_s}
\end{align}
In the second step, we have approximated the error function by taking the leading order
at $\xi(s)\ll\sqrt2\Delta$. We have also approximated the Gaussian function in the integrand
by putting $s$ to its maximum value, $s=-(1/\sqrt2)\ln(4/3)$. 
In passing, we note that $\Delta$ is assumed to be larger than the  interval of the integral, 
$(1/\sqrt2)\ln2\sim0.49$, which corresponds to the critical width 
that separates the \textit{narrow} and \textit{broad} peaks.
The integral in the second line of (\ref{small_s}) gives a constant.
Although it may be evaluated analytically, since the expression is very long and it is not illuminating,
 we do not spell it out here. Instead we simply quote its numerical value:
\begin{align}\label{Omega8}
\Omega_\text{GW,r}^{|s|<1}&
\approx0.0659\frac{\mathcal{A}_\mathcal{R}^2}{\Delta^2}\kappa^2 e^{\Delta^2}\exp\left[-\frac{\left(\ln\kappa+\Delta^2-\frac12\ln\frac43\right)^2}{\Delta^2}\right].
%\\
%&\approx0.0878\frac{\mathcal{A}_\mathcal{R}^2}{\Delta^2}\exp\left(-\frac{\ln^2\kappa}{\Delta^2}\right),
%\label{peak2}
\end{align}

\begin{figure}[htbp]
\begin{center}
\includegraphics[width=0.48\textwidth]{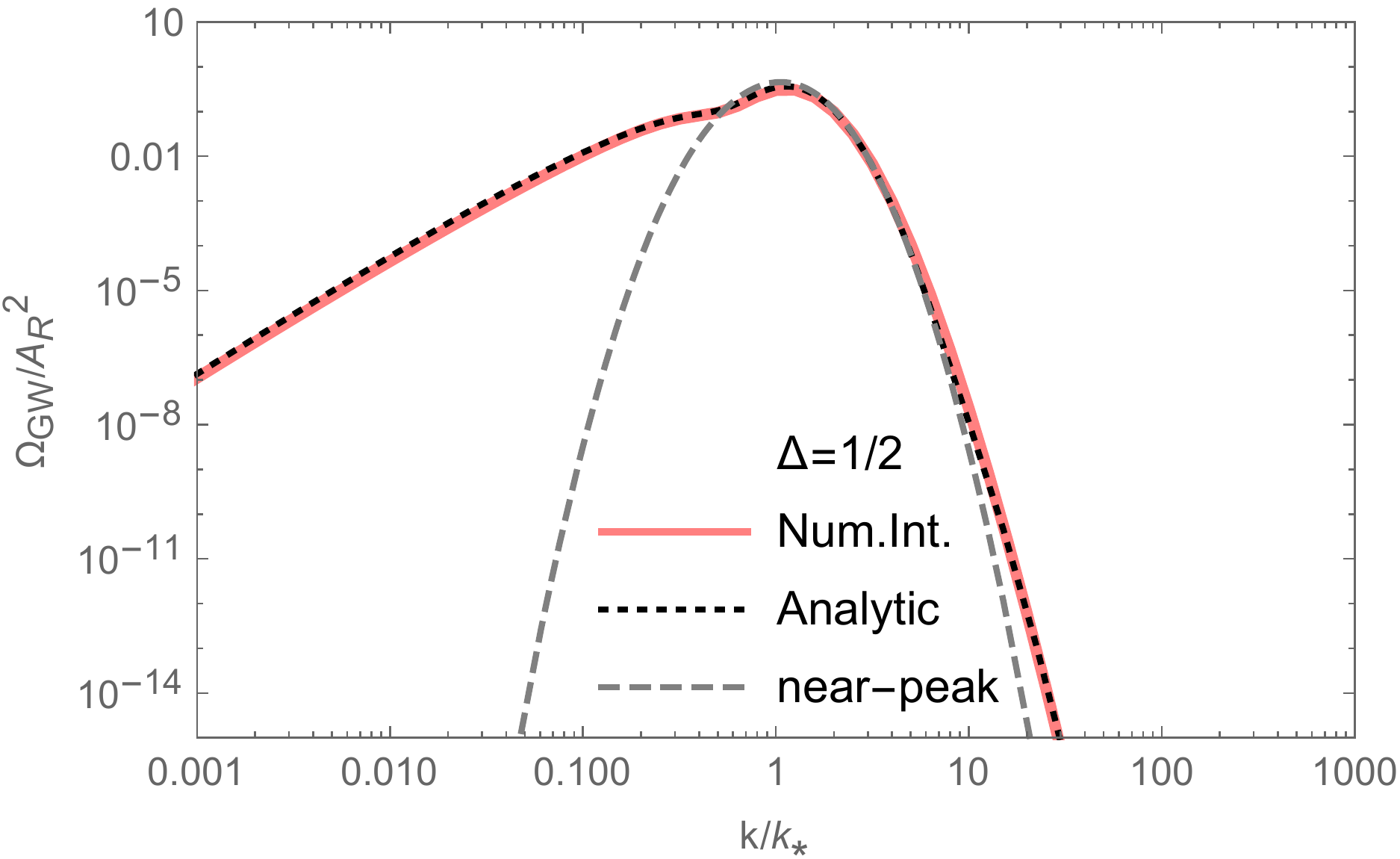}
\includegraphics[width=0.48\textwidth]{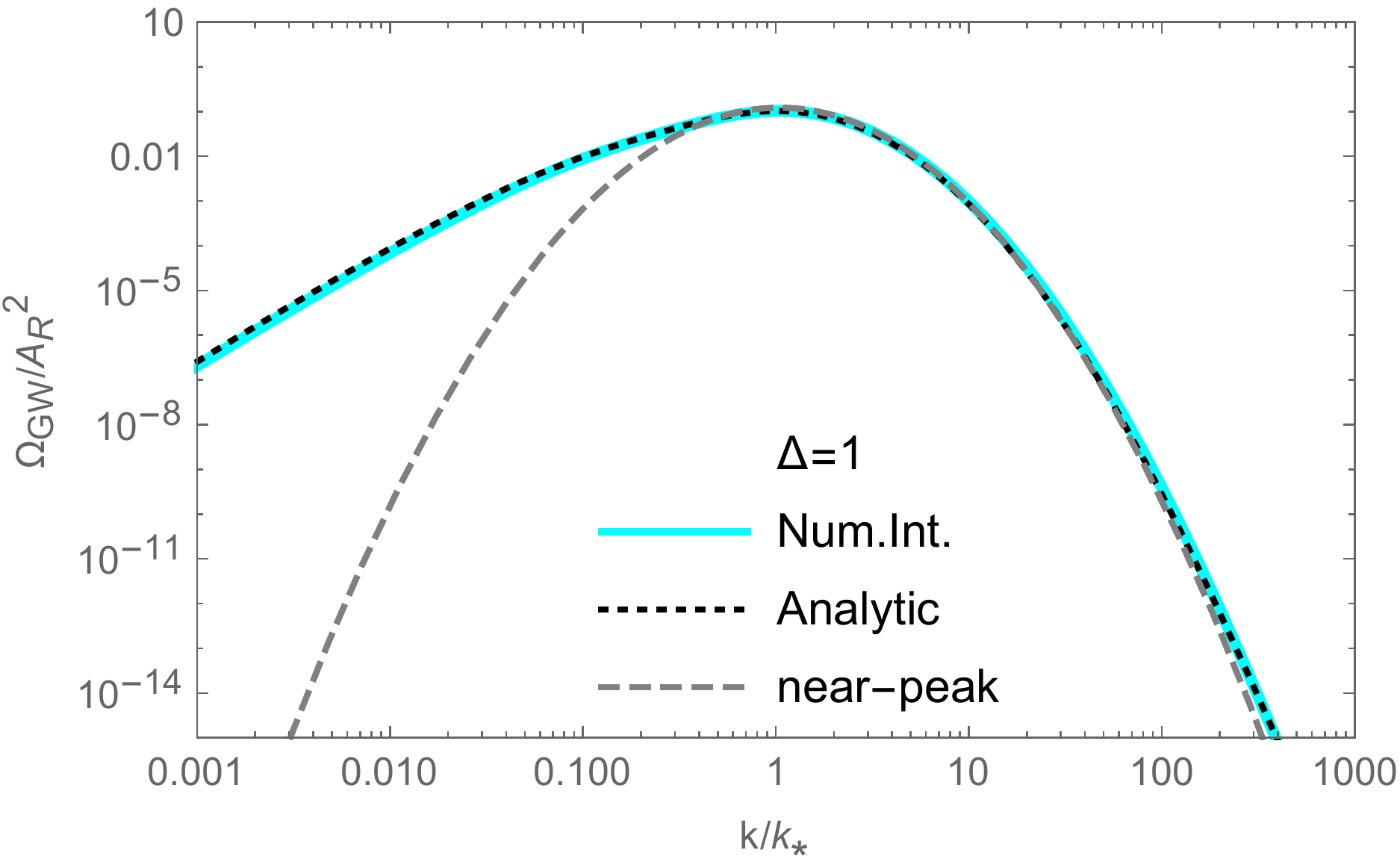}
\includegraphics[width=0.48\textwidth]{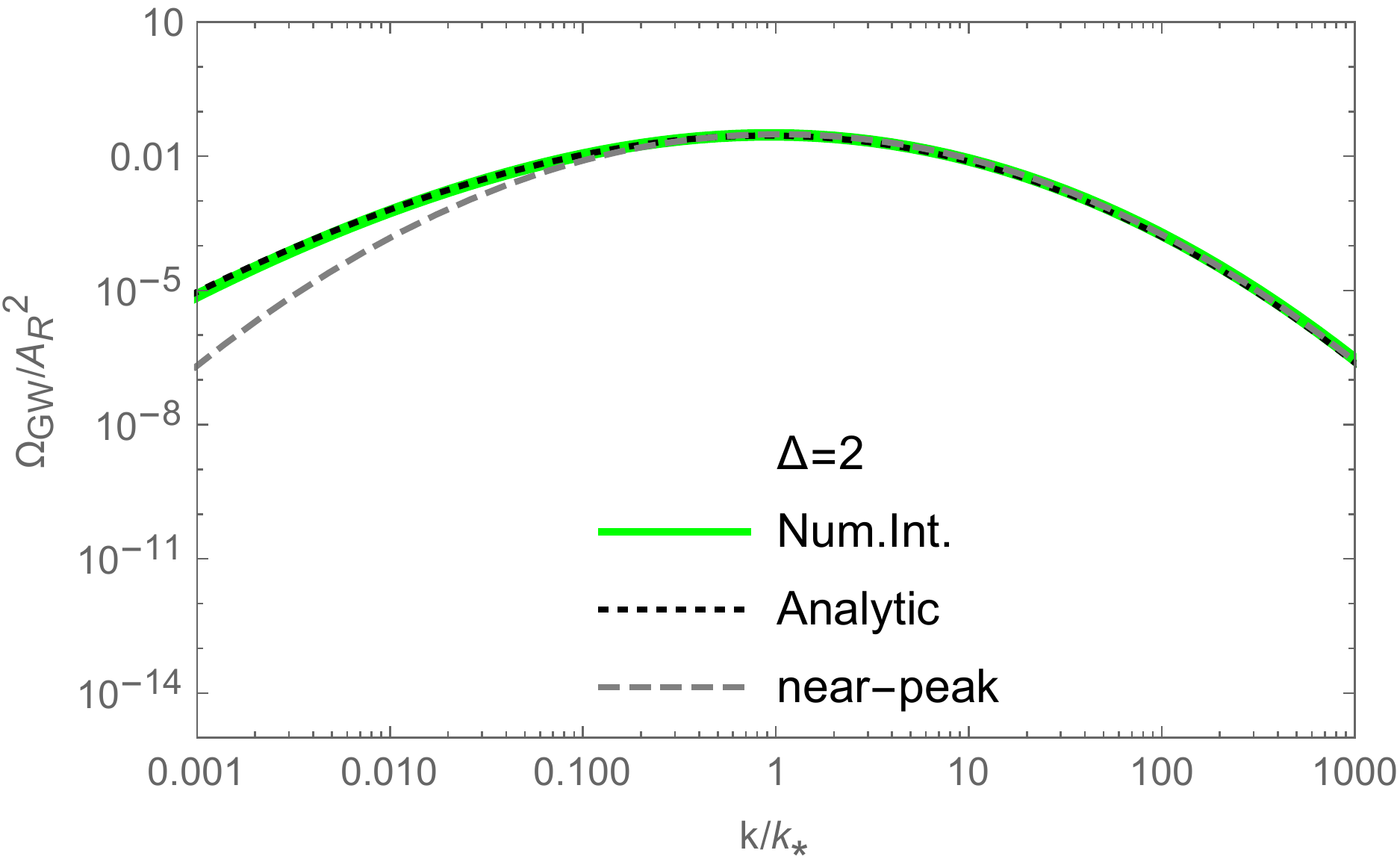}
\includegraphics[width=0.48\textwidth]{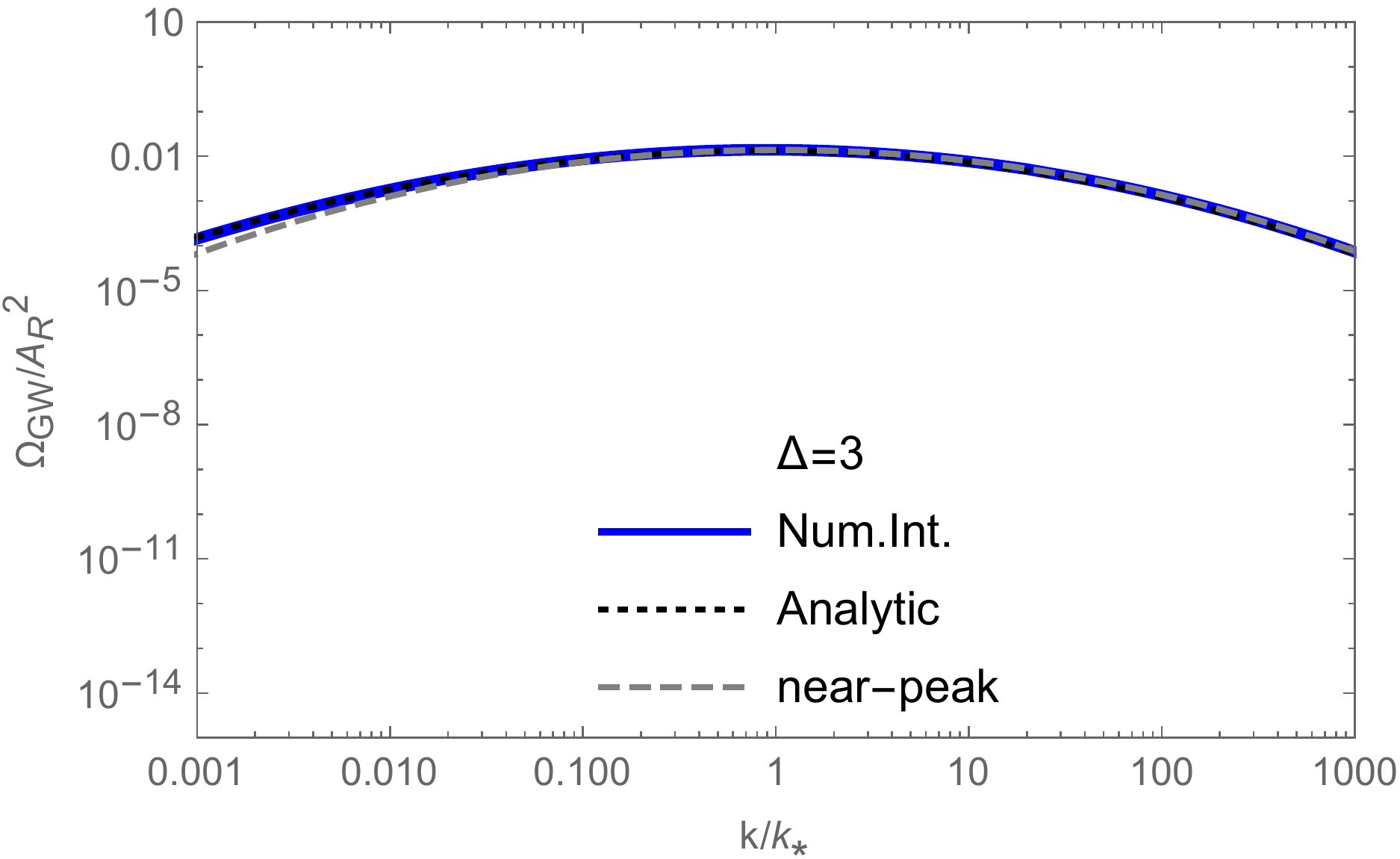}
\includegraphics[width=0.48\textwidth]{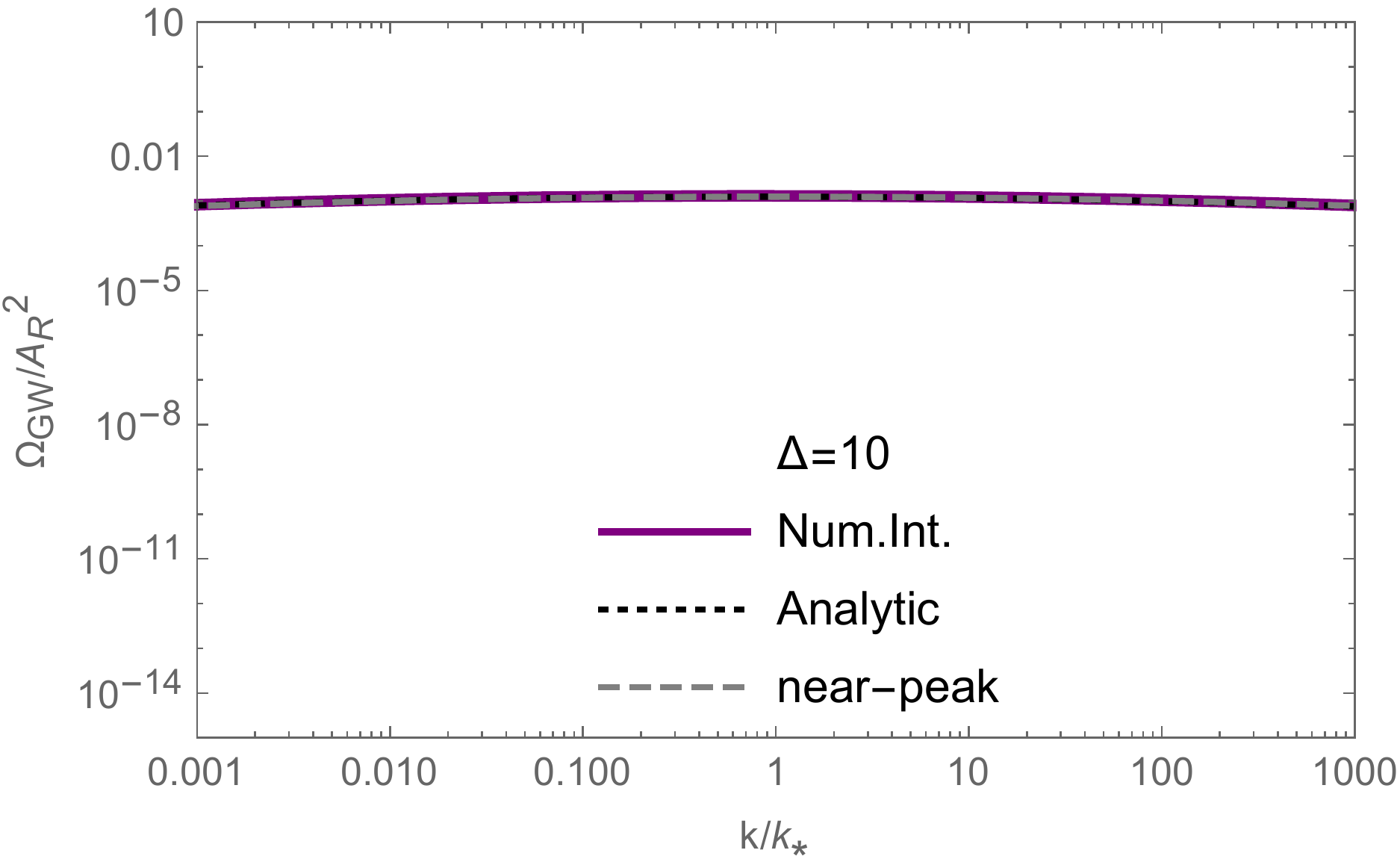}
\includegraphics[width=0.48\textwidth]{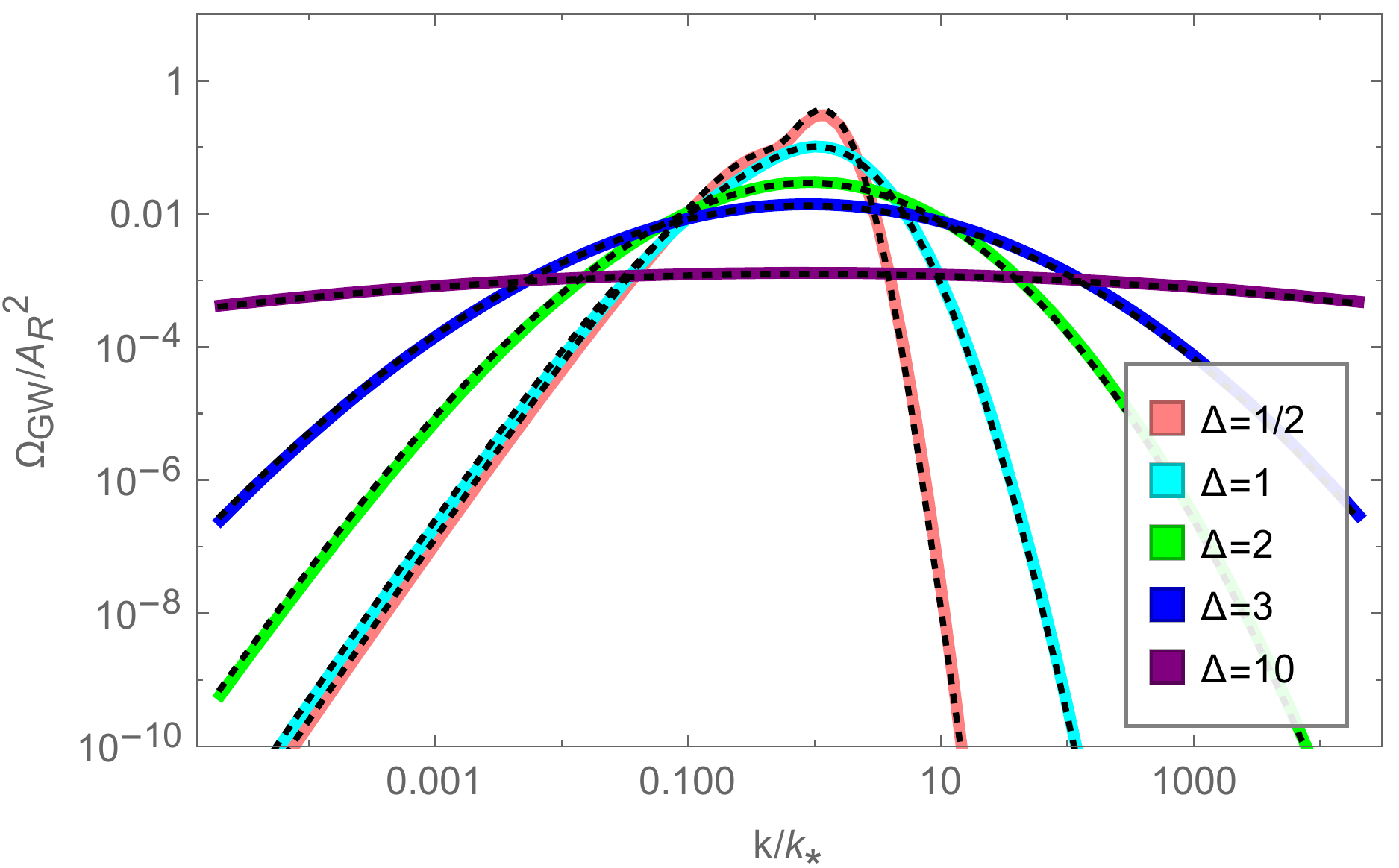}
\caption{The GW spectra for $\Delta=1/2$ (top left), $\Delta=1$ (top right), $\Delta=2$ (middle left), $\Delta=3$ (middle right), and $\Delta=10$ (bottom left), respectively. 
	 The solid thick curves are the results by numerical integration of \eqref{Omega1}.
	The black dotted curves are the approximate analytical expressions \eqref{Omega10}. 
	 The near-peak approximation \eqref{nearpeak} is also shown as gray dashed curves. 
The GW spectra for different widths are drawn together in the bottom right panel for comparison.
A  double-peak structure in the narrow-peak case is barely seen in the case $\Delta=1/2$,
which is slightly above the critical value that divides narrow and broad peaks, 
$\Delta=\Delta_c\sim 0.4$.} 
\label{fig:broad}
\end{center}
\end{figure}

For $s\lesssim-1$, the domain of integration has two strips at the leftmost region of Fig.~\ref{fig:area}. 
Along both strips, we may assume $|s|\approx|t|\gg1$,
and the leading order of $\mathscr{T}(s,t)$ is given by
\be
\mathscr{T}(s,t)=\frac{e^{4\sqrt2s}}9.
\ee
Then the integral along the $t$-axis is straightforward to find 
\begin{align}\label{temp4}
2\int^{\xi(s)}_{\chi(s)}\mathscr{T}(s,t)\exp\left(-\frac{t^2}{2\Delta^2}\right)dt
\approx\frac{\sqrt{2\pi}}9e^{4\sqrt2s}\Delta
\left(\text{erf}\left(\frac{\xi(s)}{\sqrt2\Delta}\right)-\text{erf}\left(\frac{\chi(s)}{\sqrt2\Delta}\right)\right)
\approx\frac49\sqrt2\exp\left(-\frac{s^2}{2\Delta^2}+5\sqrt2s\right).
\end{align}
Substituting \eqref{temp4} into \eqref{Omega3} and performing the Gaussian integral,
we obtain
\begin{align}\nn
\Omega_\text{GW,r}^{s\lesssim-1}
&\approx\frac{2\sqrt2}{3\pi}\mathcal{A}_\mathcal{R}^2\kappa^{-4}\frac{e^{8\Delta^2}}{\Delta^2}
\exp\left(-\frac{\ln^2\kappa}{2\Delta^2}\right)
\int^{-\sqrt2\ln2}_{-\infty}
\exp\left(-\frac{\left(s+\frac1{\sqrt2}(\ln\kappa-4\Delta^2)\right)^2}{\Delta^2}\right)ds,
\\\label{Omega9}
&=\frac1{3}\sqrt{\frac2\pi}\mathcal{A}_\mathcal{R}^2\kappa^{-4}\frac{e^{8\Delta^2}}{\Delta}
\exp\left(-\frac{\ln^2\kappa}{2\Delta^2}\right)
\text{erfc}\left(\frac{4\Delta^2-\ln(\kappa/4)}{\sqrt2\Delta}\right).
\end{align}
This expression can be further simplified by using the asymptotic behavior of the 
complementary error function. 

For $\kappa<4e^{4\Delta^2}$, which includes the near-peak region of $\kappa\sim1$,
we find
\be\label{neg1}
\Omega_\text{GW,r}^{(\text{peak},s<-1)}\approx
\frac{1}{384 \pi  \left(4 \Delta ^2-\ln (\kappa/4)\right)}\exp\left(-\frac{\ln^2\kappa+2\ln^22}{\Delta^2}\right)
\kappa^{\ln4/\Delta^2}
\approx\frac{\kappa ^{\frac{\ln (4)}{\Delta ^2}} }{1536 \pi  \Delta ^2}\exp \left(-\frac{\ln^2(\kappa)}{\Delta ^2}\right).
\ee
We see that the coefficient is much smaller than those of \eqref{peak1} and \eqref{Omega8}, 
which constitute the near-peak contributions from $s>1$ and $|s|<1$, respectively.
So we conclude that \eqref{neg1} is negligible around the peak, and reach a pretty good approximate near-peak formula as a sum of \eqref{peak1} and \eqref{Omega8},
\be\label{nearpeak}
\Omega_\text{GW,r}^\text{(near-peak)}
\approx
0.0676\mathcal{A}_\mathcal{R}^2\frac{e^{9\Delta^2/4}}{\Delta^2}
\exp\left(-\frac{\left(\ln\kappa+\frac32\Delta^2+\frac12\ln\frac32\right)^2}{\Delta^2}\right)
+0.0659\frac{\mathcal{A}_\mathcal{R}^2}{\Delta^2}\kappa^2 e^{\Delta^2}\exp\left[-\frac{\left(\ln\kappa+\Delta^2-\frac12\ln\frac43\right)^2}{\Delta^2}\right],
\ee
where we have calculated the numerical factor in \eqref{peak1}. 

For  $\kappa>4e^{4\Delta^2}$, the complementary error function gives a constant of 2, 
hence
\be\label{UV}
\Omega_\text{GW,r}^{(\text{UV},s<-1)}\approx
\frac2{3}\sqrt{\frac2\pi}\mathcal{A}_\mathcal{R}^2\kappa^{-4}\frac{e^{8\Delta^2}}{\Delta}
\exp\left(-\frac{\ln^2\kappa}{2\Delta^2}\right).
\ee
This overwhelms the contributions given by \eqref{Omega7} and \eqref{Omega8},
and gives an approximate expression of $\Omega_\text{GW,r}$ in the ultraviolet region. 
But we have to keep in mind that this ``ultraviolet'' region is far away from the peak when $\Delta\gtrsim1$,
and, because of the exponential suppression, it is probably too difficult to be detected.

To summarize, the final result of the GW spectrum when $\Delta\gtrsim\mathcal{O}(1)$ is 
the summation of \eqref{Omega7}, \eqref{Omega8}, and \eqref{Omega9}:
\begin{align}\nn
\frac{\Omega_\text{GW,r}}{\mathcal{A}_\mathcal{R}^2}
&\approx\frac{4}{5\sqrt\pi}\kappa ^3\frac{e^{\frac{9 \Delta ^2}{4}}}{\Delta}
\left[
\left(\ln^2K+\frac{\Delta^2}{2}\right)
\text{erfc}\left(\frac{\ln K+\frac12\ln\frac32}{\Delta}\right)
-\frac{\Delta}{\sqrt\pi}\exp\left(-\frac{\left(\ln K+\frac12\ln\frac32\right)^2}{\Delta^2}\right)
\left(\ln K-\frac12\ln\frac32\right)\right]\\\label{Omega10}
&+\frac{0.0659}{\Delta^2}\kappa^2 e^{\Delta^2}\exp\left(-\frac{\left(\ln\kappa+\Delta^2-\frac12\ln\frac43\right)^2}{\Delta^2}\right)
+\frac1{3}\sqrt{\frac2\pi}\kappa^{-4}\frac{e^{8\Delta^2}}{\Delta}
\exp\left(-\frac{\ln^2\kappa}{2\Delta^2}\right)
\text{erfc}\left(\frac{4\Delta^2-\ln(\kappa/4)}{\sqrt2\Delta}\right),
\end{align}
where we have introduced $K=\kappa \exp(3\Delta^2/2)$ for simplicity.
This is one of the main results of  this paper.
Its accuracy can be checked by looking at Fig.~\ref{fig:broad} where the numerical results are also shown. 
It is also useful to write down the asymptotic expressions of \eqref{IR}, \eqref{nearpeak}, 
and \eqref{UV}, each of which gives the GW spectrum valid in each different region,
\be\label{Omega11}
\frac{\Omega_\text{GW,r}}{\mathcal{A}_\mathcal{R}^2}\approx
\left\{
\begin{matrix}
\displaystyle\frac3{\sqrt\pi}\frac{e^{\frac{9 \Delta ^2}{4}}}{\Delta}
\kappa ^3\left[\left(\ln\kappa+\frac32\Delta^2\right)^2+\frac{\Delta^2}{2}\right], 
& \displaystyle\text{for}~\kappa\ll \sqrt{\frac23}e^{-\frac32\Delta^2}; & \text{(infrared)}\\
\\
\displaystyle\frac{0.125}{\Delta^2}\exp\left(-\frac{\ln^2\kappa}{\Delta^2}\right),
& \displaystyle\text{for}~\sqrt{\frac23}e^{-\frac32\Delta^2}\ll\kappa\ll4e^{4\Delta^2}; &\text{(near peak)}\\
\\
\displaystyle\frac2{3}\sqrt{\frac2\pi}\kappa^{-4}\frac{e^{8\Delta^2}}{\Delta}
\exp\left(-\frac{\ln^2\kappa}{2\Delta^2}\right),
& \text{for}~\kappa\gg4e^{4\Delta^2}. &\text{(ultraviolet)}\\
\end{matrix}
\right.
\ee
Although each of the above asymptotic formula fits very well in each region,
we only plot the most important, near-peak formula in Fig.~\ref{fig:broad},
which is \eqref{nearpeak}. When $\Delta\gtrsim1$, it can be further simplified as
\be\label{sumpeak}
\Omega_\text{GW,r}^\text{(peak)}
\approx0.125\frac{\mathcal{A}_\mathcal{R}^2}{\Delta^2}
\exp\left(-\frac{\ln^2\kappa}{\Delta^2}\right).
\ee
This formula for the near-peak region is another main result of this paper. 
It clearly shows that the induced GW spectrum has a lognormal peak with width $\Delta/\sqrt2$
which is smaller by $\sqrt{2}$ of the peak width of the original curvature perturbation spectrum, $\Delta$. 
This is a reflection of the ``secondary'' nature of the induced GWs, as $\Omega_\text{GW,r}\propto(\mathcal{P}_\mathcal{R})^2$.\footnote[5]{We thank 
	Rong-Gen Cai for pointing out this simple physical explanation for the relation between the two widths.}
The maximum of the GW spectrum is given directly by setting $\kappa=1$, 
\be\label{broadmax}
\Omega_\text{GW,r}^\text{(max)}
%\approx\frac{8+12\ln\frac32+9\ln^2\frac32}{36\pi}\frac{\mathcal{A}_\mathcal{R}^2}{\Delta^2}
\approx0.125\frac{\mathcal{A}_\mathcal{R}^2}{\Delta^2}.
\ee
 Since (\ref{sumpeak}) gives the height and width of the peak of the induced GW spectrum, 
we expect it to play an important role in the interpretaion of observational signals in the near future.

An important special case of a broad peak is the GWs induced by a scale-invariant curvature perturbation, $\mathcal{P}_\mathcal{R}(k)=A_\mathcal{R}$. 
Comparing to \eqref{lognormalpeak}, this corresponds to the limits
$\mathcal{A}_\mathcal{R}\rightarrow\infty$ and $\Delta\rightarrow\infty$ with the ratio  
$\mathcal{A}_\mathcal{R}/(\sqrt{2\pi}\Delta)\equiv A_\mathcal{R}=\text{const.}$. 
This gives a scale-invariant GW spectrum,
\be
\Omega_\text{GW,r}=0.783A_\mathcal{R}^2,
\ee
which is in good agreement with the result obtained numerically in Ref.~\cite{Kohri:2018awv}.

\section{Conclusion}\label{sec:con}
In this expanding new era of GW astronomy/cosmology, detecting the stochastic background of GWs
is the next scientific goal for the GW experiments in the coming decades.
Among possible sources of stochastic backgrounds, the secondary GWs induced by the primordial 
curvature perturbation is a very important target. As both the generation of GWs 
and the formation of PBHs occur essentially at the horizon reentry of the relevant scale in the early universe,
the fruitful PBH physics motivated us to study the associated induced GWs.
Observations on both sides can be used to probe the primordial curvature perturbation on small scales, 
where there is no stringent constraints. 

In this paper we have studied the spectrum of GWs induced by the primordial curvature perturbation 
whose spectrum has a lognormal peak.
 Such a spectrum is frequently discussed in the literature from the viewpoints of
both inflationary phenomenology and PBH formation mechanism. 
We have found that the resulting spectrum of the induced GWs has
distinct features depending on the width of the longnormal spectrum, $\Delta$.
For both narrow-peak ($\Delta\ll1$) and broad-peak ($\Delta\gtrsim1$) cases, 
we have successfuly obtained an analytical formula with amazingly good accuracy,
as presented in \eqref{Omega4} for $\Delta\ll1$ and \eqref{Omega10} for $\Delta\gtrsim1$,
respectively. 

For the narrow-peak case, the GW spectrum \eqref{Omega4} can be further simplified
 by using the GW spectrum for a $\delta$-function peak. Namely, it may be factorized
 into two components,
\be\label{Omega12}
\Omega_{\text{GW,r}}^{(\Delta\ll1)}\approx
\text{erf}\left(\frac1\Delta\text{arcsinh}\frac{k}{2k_*}\right)\Omega_{\text{GW,r}}^{(\delta)},
\ee
where the factor given by the error function is independent of the equation of state of the universe,
and $\Omega_{\text{GW,r}}^{(\delta)}$ is for the $\delta$-function curvature perturbation spectrum
whose formula in the radiation-dominated universe, \eqref{Omega2}, is well known,
and it is not difficult to extend it to that for a universe with arbitrary equation of state. 
Thus (\ref{Omega12}) gives a universal formula valid for any equation of state of the universe if
$\Delta\ll1$. 

 As $\Omega_{\text{GW,r}}^{(\delta)}\propto k^2$ in the radiation dominated universe,
  \eqref{Omega12} clearly shows there is a break in the power-law at $k_b\approx2k_*\Delta$ 
  on the infrared side of the GW spectrum, changing from $k^3$ to $k^2$ as $k$ increases.
  In addition, there appears a logarithmically diverging peak at $k_p\approx 2k_*/\sqrt{3}$.
  If such a GW spectrum is observed in the future, with both break and peak frequencies measured, 
  the width of the peak in the original curvature perturbation  spectrum can be read off as
\be
\Delta\approx\frac{f_b}{\sqrt{3}f_p}\,,
\ee
where $f_b$ and $f_p$ are the break and peak frequencies, respectively.

For the broad-peak case, we have also derived an analytical expression \eqref{Omega10} which fits 
 the numerical result very well. 
 Further, we have obtained the asymptotic expressions for infrared, near-peak, and ultraviolet regions,
  respectively.
  The most important feature of the GW spectrum for the broad-peak case is
the appearance of  a lognormal peak with width $\Delta/\sqrt2$, which is smaller by $\sqrt{2}$
compared with the width $\Delta$ of the primordial curvature perturbation, as shown in \eqref{sumpeak},
which reflects the secondary nature of the induced GWs. 

The maximum of the GW spectrum has been also derived, 
\be
\frac{\Omega_\text{GW,r}^\text{(max)}}{\mathcal{A}_\mathcal{R}^2}
=\begin{dcases}
\displaystyle\frac49\Big(\ln(2\Delta)+1\Big)^2+2.64 & \text{for}~\delta f<\Delta\ll1,\\
\\
\displaystyle\frac{0.125}{\Delta^2} & \text{for}~\Delta\gtrsim1.
\end{dcases}
\ee
As we stated at the end of Section \ref{sec:setup}, all the results of $\Omega_\text{GW,r}$ listed above 
are valid only until the epoch of matter-radiation equality. To obtain the GW spectrum we observe
today, they should be multiplied by a redshift factor of $2\Omega_{r,0}\sim8.2\times10^{-5}$.

In this paper we have only considered lognormal spectra for the primordial curvature perturbation.
Such spectra may arise in varieties of inflation models, which implies the usefulness of our results.
Nevertheless, the primordial curvature perturbation may have a more complicated spectrum. 
For example, the spectrum may be composed of  Gaussian fluctuations and non-Gaussian corrections. 
In such a case, even if the Gaussian part has a lognormal spectrum, the non-Gaussian part
may substantially deform the spectral shape if its contribution is large.
A specific example was discussed in Ref.~\cite{Cai:2018dig} where the local-type non-Gaussianity
is added on top of the Gaussian perturbation: 
$\calR=\calR_g+F_{\rm NL}\left(\calR_g^2-\left\langle\calR_g^2\right\rangle\right)$ 
with a narrowly peaked spectrum of $\calR_g$.
%In this case the curvature perturbation spectrum is no longer narrowly peaked, and it was found that there appears another peak locating at $(4/\sqrt3)k_*$, with width $2\Delta$ and amplitude of $\mathcal{O} \left((F_\text{NL}\mathcal{A}_\mathcal{R})^4\right)$. 
Deriving analytical expressions for more general forms of the primordial curvature perturbation
spectrum is a challenging issue left for future work.

\begin{acknowledgements}
We thank Rong-Gen Cai and Guillem Dom\`enech for useful discussions. 
S.P. especially thanks his father Tian-Ming Pi for taking care of him when he was working in his hometown 
under the lockdown in Hubei Province, China during the COVID-19 pandemic. 
The work of S.P. is supported in part by JSPS Grant-in-Aid for Early-Career Scientists No. 20K14461. 
The work of M.S. is supported in part by the JSPS KAKENHI Nos. 19H01895 and 20H04727. 
Both S.P. and M.S. are supported by the World Premier International Research Center Initiative
 (WPI Initiative), MEXT, Japan. 
\end{acknowledgements}

\end{document}